\documentstyle[11pt,newpasp,twoside,epsf]{article}

\def\edcomment#1{\iffalse\marginpar{\raggedright\sl#1\/}\else\relax\fi}
\reversemarginpar

\begin{document}
\title{Determination of Substellar Mass Function of Young Open Clusters Using
2MASS and GSC Data}
\author{Anandmayee Tej}
\affil{Observatoire Astronomique de Strasbourg, 67000 Strasbourg, France}
\author{Kailash C. Sahu}
\affil{Space Telescope Science Institute, Baltimore, MD 21218, USA}
\author{T. Chandrasekhar}
\affil{Physical Research Laboratory, Navrangpura, Ahmedabad - 380009, India}
\author{N. M. Ashok}
\affil{Physical Research Laboratory, Navrangpura, Ahmedabad - 380009, India}

\begin{abstract}
We present a statistical method to derive the mass functions of open
clusters using sky survey data such as the 2 Micron All Sky Survey
(2MASS) and the Guide Star Catalogue (GSC). We have used this method to
derive the mass functions in the stellar/substellar regime of three
young, nearby open clusters, namely IC 348, $\sigma$ Orionis and
Pleiades. The mass function in the low mass range (M$<0.50 M_\odot$) is
appreciably flatter than the stellar Salpeter function for all three
open clusters. The contribution of objects below 0.5~M$_\odot$ to the
total mass of the cluster is $\sim$40\% and the contribution of objects
below 0.08~M$_\odot$ to the total is $\sim$4\%.

\end{abstract}

\section{Introduction}
Recent surveys have found a significant population of low-mass stars,
brown dwarfs and planetary-mass objects in young open clusters. Since
low-mass objects evolve little over the lifetime of the Universe, the
present day mass function of these objects is a good representation of
the Initial Mass Function (IMF). The mass function in this low-mass
regime is however poorly known due to faintness of these objects and
also due to uncertainty in the mass-luminosity relations. Low-mass
objects at or below the Hydrogen Burning Mass Limit (HBML) of
0.08~M$_\odot$ are known to be warmer and hence more luminous  when
young although they cool rapidly and fade with age (Baraffe et al.
1998). The combination of youth and proximity in some open clusters
make them ideal targets for searches of low-mass objects below the HBML
particularly at infrared wavelengths. In the present study we have
adopted a statistical approach to determine the mass function (dN/dM
$\propto$ M$^{-\alpha}$) of objects in the mass range $0.5~M_\odot$ to
0.025--0.05~M$_\odot$ using data of three open clusters namely IC 348,
$\sigma$ Orionis and Pleiades. 

\section{Sample Selection and Analysis} 
We have used the data from two extended sky surveys --- 2MASS Second
Incremental Release and the latest version of GSC --- on the three open
clusters.  The limiting magnitudes of 16.5, 15.5, 15 and 21 in the $J$,
$H$, $K$ and $F$ (POSS II IIIa-F) passbands, respectively,  enable us
to probe down to 0.025~M$_\odot$  in IC 348 and $\sigma$ Orionis and
0.04M$_\odot$ in Pleiades. Unlike most other previous studies which
rely on confirming candidate low-mass objects through spectroscopy we
use a statistical approach to estimate the number of low-mass objects.
In this approach it is important to use several control fields close to
each cluster to subtract the contribution of foreground and background
objects. The nature of these two extended surveys permitted us to use
several control fields and variable field sizes. Table 1  lists the
positions and radii of the fields chosen for the three clusters.

\begin{table}
\caption{Positions and sizes of the fields}

\begin{tabular}{lllcl}
\\
\tableline
Fields&RA (J2000)&Dec (J2000)&Radius\\
&(h m \ s)&( $^\circ$ \ \  \ $\arcmin$ \ \ $\arcsec$)&(arcmin)\\
\tableline
IC 348&03 44 30&+32 17 00&20\\
Control 1&03 49 08&+31 19 08&20\\
Control 2&03 44 10&+33 19 26&20\\
&\\
$\sigma$ Orionis&05 38 45&--02 36 00&30\\
Control 1&05 58 29&--04 29 48&30\\
Control 2&0.5 11 00&--00 20 00&30\\
&\\
Pleiades&03 47 00&+24 07 00&90\\
Control 1&03 18 00&+26 41 00&90\\
Control 2&03 05 00&+24 42 00&90\\
\tableline
\end{tabular}
\end{table}

In the first stage of analysis we merged the 2MASS and the GSC sources
by taking 2MASS coordinates and cross correlating them with the GSC
catalogue. A search radius of 2\arcsec was found to be the optimum
value for the cross correlation being small enough to reject spurious
and multiple detections and large enough to include any minor
positional uncertainties in the two catalogues. Selection criteria to
pick out cluster members were derived by inspecting various colour
magnitude diagrams (CMDs). The use of the F versus (F--R) CMD overcomes
the degeneracies between the mass and  distance found in the K$_s$
versus (J--K$_s$) CMD, and minimizes the overlap between reddened
background stars and low-mass members of the cluster. This leads to a
more efficient rejection criteria for non members which is accomplished
using the empirical data from Leggett (1992) and the theoretical
isochrones of Baraffe et al. (1998). The selection criteria for
low-mass members, derived using the suitably distance-scaled and
extinction corrected model isochrones of Baraffe et al. (1998), 
are listed in Table 2. The sources satisfying the criteria are further
statistically filtered for possible contaminants using the control
fields. Details of the procedure are discussed in Tej et al. (2002).
Finally, the masses of the selected candidates are determined by
comparing observed magnitudes with those derived from evolutionary
models (Baraffe et al. 1998 and Chabrier et al. 2000).

\begin{table}
\caption{Criteria for low-mass cluster members}
\begin{tabular}{llll}
\\
\tableline
IC 348&$\sigma$ Orionis&Pleiades\\
\tableline
$F<[2.58(F-J)$&$F<[2.55(F-J)$&$F<[7.93+1.99(F-J)$\\
\hspace*{0.65cm} +10.14]&\hspace*{0.65cm} +10.26]&$\hspace*{0.65cm} +056(F-J)^2-0.13(F-J)^3$\\
&&$\hspace*{0.65cm} +0.01(F-J)^4]$\\
$F-K \geq 4.09$&$F-K \geq 4.35$&$F-K \geq 3.42$\\
$K \hspace*{0.75cm} \geq 11.20$& $K \hspace*{0.75cm}\geq 11.17$&
$K \hspace*{0.75cm} \geq 11.17$\\
$J-K \geq$ 0.90&$J-K \geq$ 0.94&$J-K \geq$ 0.83\\
$H-K \geq$ 0.20&$H-K \geq$ 0.20&$H-K \geq$ 0.22\\
\tableline
\end{tabular}
\end{table}

\begin{figure}[ht]
\centering \leavevmode
\epsfxsize=10truecm \epsfbox{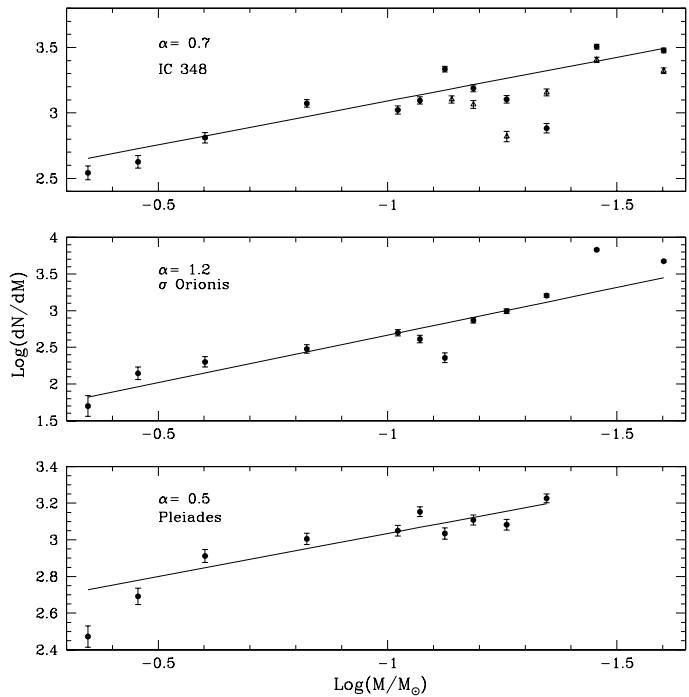}
\caption{\small{The derived mass functions for the three clusters. The
y-error bars are the $\sqrt{N}$ errors involved in the counting
statistics. For all three clusters the filled circles are points
derived using the models of Baraffe et al. (1998) and for IC 348
the open triangles are data points derived using the dusty
isochrones of Chabrier et al. (2000).}} 
\end{figure}

\section{Results}

We first carried out a detailed study for IC 348. For this cluster, we
derived the mass function using the solar metallicity models of Baraffe
et al (1998) and dusty models of Chabrier et al. (2000) both of which
gave similar results. The location of low-mass members isolated by us
and the spectroscopically confirmed low-mass members of Luhman (1999)
are in good agreement which builds the confidence in our selection
criteria. We then used the same methodology for $\sigma$ Orionis and
Pleiades. The resulting exponents of the mass functions for IC 348,
$\sigma$ Orionis and Pleiades are respectively 0.7, 1.2 and 0.5 with an
estimated error of $\pm$ 0.2. For the younger clusters IC 348 and
$\sigma$ Orionis, there is some hint of a possible dip in the mass
function at the position of the HBML.

\begin{table}
\caption{Summary of the results}
\begin{tabular}{llllll}
\\
\tableline
Cluster&Age&Distance&Mass Range&$\alpha$\\
&(Myr)&(pc)&(M$_\odot$)\\
\tableline
IC 348&5&316&0.5--0.035&0.7\\
$\sigma$ Orionis&3&352&0.5--0.045&1.2\\
Pleiades&100&125&0.5--0.055&0.5\\
\tableline
\end{tabular}
\end{table}

The results of this statistical approach (Table 3) imply that though
the mass function continues to rise above the HBML it is appreciably
flatter compared to the Salpeter mass function. In Figure 1 we have
plotted the derived mass function for the three clusters. Taking the
canonical Salpeter exponent of 2.35 in the mass range 1--10 M$_\odot$,
the Chabrier exponent of 1.55 in the mass range 0.5--1.0 M$_\odot$ and
the values obtained by us below 0.5 M$_\odot$ we estimate the mass
contribution to be about 40\% for objects below 0.5 M$_\odot$ and 4\%
for objects below the HBML of 0.08 M$_\odot$. Our results are
consistent with previous studies (e.g. B\'ejar et al., 2001) and suggest
that although low-mass objects are at least as numerous as their
stellar counterparts their contribution to the total mass of the
cluster is small. 

\acknowledgments
We would like to thank Brian McLean and Mario Lattanzi (GSC-II project
scientists) for access to the development version of GSC-II in advance
of publication. We are grateful to I. Baraffe and F. Allard for making
electronic versions of the latest models available and generating model
isochrones for the nonstandard F passbands.

\end{document}